\documentclass[a4paper,fleqn]{cas-sc}

\usepackage[numbers]{natbib}
 
\usepackage{tikz}
\usepackage{float}
\usetikzlibrary{arrows}
\usepackage{units}
\usepackage{setspace}
\usepackage{tabularx}
\usepackage{booktabs}
\usepackage{subfig}
\usepackage{todonotes}
\usepackage{nicefrac}
\usepackage{color}
\usepackage{upgreek}

\def\reff#1{\mbox{\rm(\ref{#1})}}

\definecolor{orange}{RGB}{255,127,0}
\definecolor{greeen}{rgb}{0.12, 0.3, 0.17}
\definecolor{dgreen}{RGB}{34,139,34}
\definecolor{dblue}{RGB}{10,25,150}

\begin{document}
\let\WriteBookmarks\relax
\def\floatpagepagefraction{1}
\def\textpagefraction{.001}

\shorttitle{High-resolution IC for measuring Radon decay in air}

\shortauthors{R. Nötzel, K. Weinberg}

\title[mode = title]{Design of a high-resolution ion pulse ionization chamber for ${^{222}}$Rn decays detection in air}

\author[1]{Ralf Nötzel}  
\ead{ralf.noetzel@uni-siegen.de}

\credit{Writing – first draft, investigation, chamber construction, measurement,  software, visualization}

\author[1]{Kerstin Weinberg}[orcid=0000-0002-2213-8401]
\cormark[1]
\ead{kerstin.weinberg@uni-siegen.de}
\ead[URL]{https://www.mb.uni-siegen.de/fkm/index.html.en?lang=en}
\credit{Writing – draft, review and editing, conceptualization, formal analysis, supervision}

\affiliation[1]{organization={Department of Mechanical Engineering, Festkörpermechanik, Universität
Siegen},
            addressline={Paul-Bonatz-Straße 9-11},
            city={Siegen},
            citysep={},
            postcode={57076},
            state={NRW},
            country={Germany}}


\begin{abstract}
Radon is a naturally occurring radioactive gas that contributes significantly to human radiation exposure and must be controlled to avoid concentrations harmful to health.

The paper presents an impulse-proportional ionization chamber that is suitable for the direct measurement of low radon concentrations in ambient air and achieves spectroscopic resolutions of 2-3\%.

This accuracy can be achieved through two novel principles. On the one hand, the IC's double-cylindrical, coaxial design allows for efficient, nearly complete detection of $\alpha$-radiation. Second, a customized measurement method for spectroscopic evaluation was developed to discriminate between the proportions of {${^{222}\mathrm{Rn}}$, ${^{218}\mathrm{Po}}$, and ${^{214}\mathrm{Po}}$} and extract their concentrations. Particular attention was paid to the energy resolution of the detection system by suppressing the effects of acoustic and vibration noise on the detector's operation.

The high spectral resolution of the developed ionization chamber, with working volumes of 7.7 and 8.7 l, enables measurements with uncertainties of less than 5\% at 15-minute measurement times in ambient air with $50\,\mathrm{Bq/m^3}$ radon activity.
\end{abstract}





\maketitle
\section{Motivation}\label{sec:Intro}

Radon is a colorless, odorless, and chemically inert noble gas that is radioactive and decays while emitting $\alpha$ radiation. It contributes significantly to human radiation   exposure. Radon is primarily released in geological formations containing uranium and radium, particularly in granite, slate, claystone, and volcanic rocks. In outdoor air, it quickly dilutes to typical activity concentrations of a few $\mathrm{Bq/m^3}$. However, in closed rooms, basements, disused mine tunnels, or groundwater wells, a significant accumulation can occur, with values of several $100\,000\, \mathrm{Bq/m^3}$.

When radon-containing air is inhaled, a certain proportion decays into solid, $\alpha$ emitting decay products that remain in the lungs. There, it damages the tissue. Radon is therefore the second most common cause of lung cancer after smoking. Although the phenomenon was described as miners' disease by Paracelsus and the gas was detected as early as 1900, legal regulations for protection against radon have only been in place in the EU since the 2010s \cite{BfS2024,EU}. Of particular importance here is ${^{222}\mathrm{Rn}}$, which has a half-life of 3.82 days, whereas ${^{220}\mathrm{Rn}}$ (thoron) decays after only 54 seconds.
The main elements in the decay series belonging to ${^{222}\mathrm{Rn}}$ are: \begin{align*}
{^{226}\mathrm{Ra}} \xrightarrow{\alpha}\,
{^{222}\mathrm{Rn}} \xrightarrow{\alpha}\,
{^{218}\mathrm{Po}} \xrightarrow{\alpha} \cdots\,
{^{214}\mathrm{Po}} \xrightarrow{\alpha} \cdots\,
{^{210}\mathrm{Po}} \xrightarrow{\alpha}\,
{^{206}\mathrm{Pb}} \quad (\mathrm{stabil})
\end{align*}

In addition, radon concentrations in ambient air are also of interest in physical experiments and weather observations. In recent years, for example, studies have shown elevated radon levels shortly before earthquakes \cite{Chambers2018}.

In all these cases, it is necessary to measure radon concentration as accurately and quickly as possible. Three different methods have been established for this purpose:

The \emph{Lucas cell} with a photomultiplier (PMT) is a classic method for detecting $\alpha$ particles \cite{LukasWoodward1964}, widely used in many professional devices. It does not measure radon itself, but rather the flashes of light (scintillations) 
of the $\alpha$ particles from ${^{222}\mathrm{Rn}}$ and its decay products on a ZnS(Ag) scintillator layer. 
The measurement method has been technically optimized and enables time-resolved measurements of ${^{222}\mathrm{Rn}}$ concentrations to below 1\,Bq/m$^3$
\cite{WilliamsChambers2016_HistoryColl,Sarad2024}. 
A disadvantage is the considerable size of the systems when designed for extremely low radon concentrations. Due to the nature of the system, spectroscopy is not possible; only the decay products of ${^{222}\mathrm{Rn}}$ are detected.

\emph{PIN semiconductor diodes} do not measure radon gas directly, but instead detect the $\alpha$ particles of the nuclides  ${^{218}\mathrm{Po}}$ and ${^{214}\mathrm{Po}}$ and, if applicable, 210Po. A measuring chamber containing 0.5–2 liters of ambient air is typically used for this purpose.   Each $\alpha$ decay produces an energy-dependent voltage pulse at the PIN diode surface. Because the effective measuring surface is only a few square centimeters, the decay products are collected using an electric field \cite{Pronost_etal2019}.
The method enables accurate spectroscopic evaluation and is primarily used in the scientific field.

Ionization chambers (ICs), especially pulse-proportional ICs with airflow, enable a direct measurement of ${^{222}\mathrm{Rn}}$.
The $\alpha$ particles of the decaying radon and its decay products ionize the air, creating free charge carriers that are collected by an electric field.
The concentration of ${^{222}\mathrm{Rn}}$ is determined from the resulting charge pulse. Radon measurement with IC is precise and therefore used in calibration laboratories, for long-term monitoring, and in research, see e.g.
\cite{Kuzminov2003,Gavrilyuk_etal2011,WilliamsChambers2016_HistoryColl,Kozlov_etmany2020,Etezov_etal2022}.

Regardless of the measurement principle, as much air as possible should be used for the measurement to achieve an acceptable measurement uncertainty with the shortest possible measurement time. For example, at a radon concentration of $100\,\mathrm{Bq/m^3}$ in a $1 $l measuring chamber, $n=180$ decays of ${^{222}\mathrm{Rn}}$ occur within 30 minutes. With the statistical assumptions of a Poisson distribution, the relative uncertainty of the measurement is $1/\sqrt{n}$, i.e., 7.5\%. With a measurement volume of $5 $l, this uncertainty is already reduced to 3\%.

In practice, however, existing ICs have a significantly higher measurement error. Therefore, this article presents a measuring chamber based on the ideas of Gavrilyuk et al. \cite{Gavrilyuk_etal2015} that enables significantly more accurate radon measurements in ambient air. The design of this IC combines high-resolution spectroscopy with a large chamber volume, resulting in short measurement times.

The article is structured as follows: First, Section \ref{sec:IIC} discusses typical IC designs and presents the proposed cylindrical $\alpha$-pulse proportional ionization chamber construction. Measurement setup and data evaluation are outlined and verified in Section \ref{sec:Elektrik}. In Section \ref{sec:Results}, we  exemplary show measurements results of ${^{222}\mathrm{Rn}}$ and its decay products in ambient air, achieving accuracies of 2-3\% . Section \ref{sec:Summary}  summarizes our results briefly.

\section{Design of the impulse ionisation chamber}\label{sec:IIC}

The basic design of ionization chambers is always similar. Two electrodes form a capacitor with air as the dielectric. Radon and its decay products ionize the air in the IC through the emitted $\alpha$ particles, producing $\mathrm{O^{2+}}$ and $\mathrm{N^{2+}}$ ions as well as electrons.
The electrons recombine with oxygen within microseconds,
while the ions move toward the oppositely charged electrodes. At typical voltages of $1\dots 2\,\mathrm{kV}$, the ions drift about $5\dots 10\,$mm/ms.
The collected charges are measured as voltage pulses.
\begin{figure}
    \centering
    \includegraphics[width=0.45\linewidth]{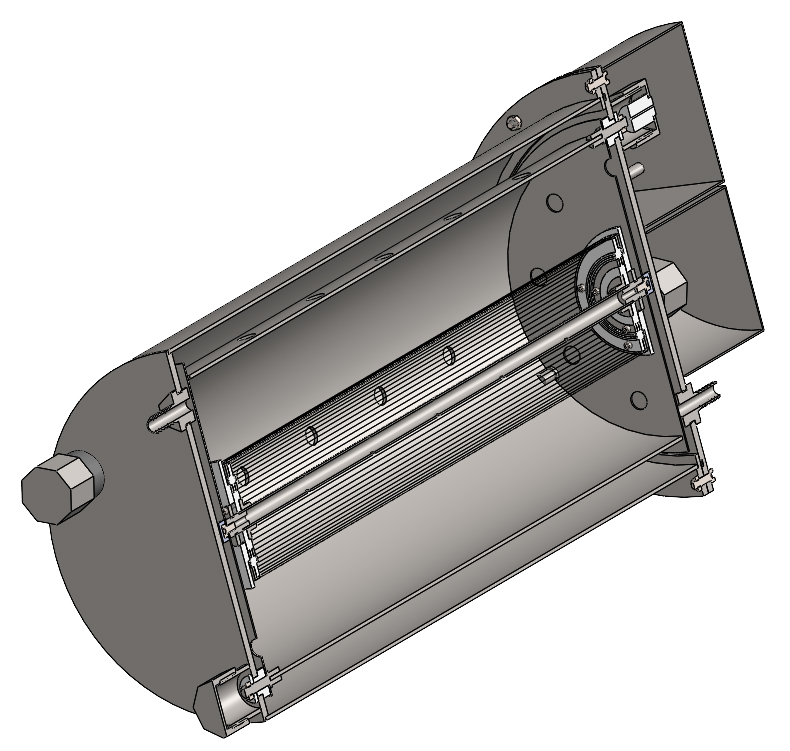}
    \caption{CAD drawing showing a cut-through of the cylindrical impulse-proportional ionisation chamber (CIPIC I)}
    \label{fig:CIPICcut}
\end{figure}

Two design principles are evident for the construction of ICs:
\begin{itemize}
  \item[\textbf{I}] The IC is constructed in such a way that the electric field and the path lengths to be bridged by the ions inside the chamber are kept as constant as possible.   As a result, the maximum value of the detected signal is proportional to the energy of the incident $\alpha$ particle and can be evaluated using spectroscopy.
  \item [\textbf{II}] The IC is constructed in such a way that the $\alpha$ particles can fly as far as possible without colliding.   As a result, all $\alpha$ particles generate ions, but they take different times to reach the electrode.      Here, the integral of the detected signal over time is proportional to the energy of ${^{222}\mathrm{Rn}}$ and decay products.
\end{itemize}

\begin{figure}
    \centering
    \includegraphics[height=0.42\linewidth]{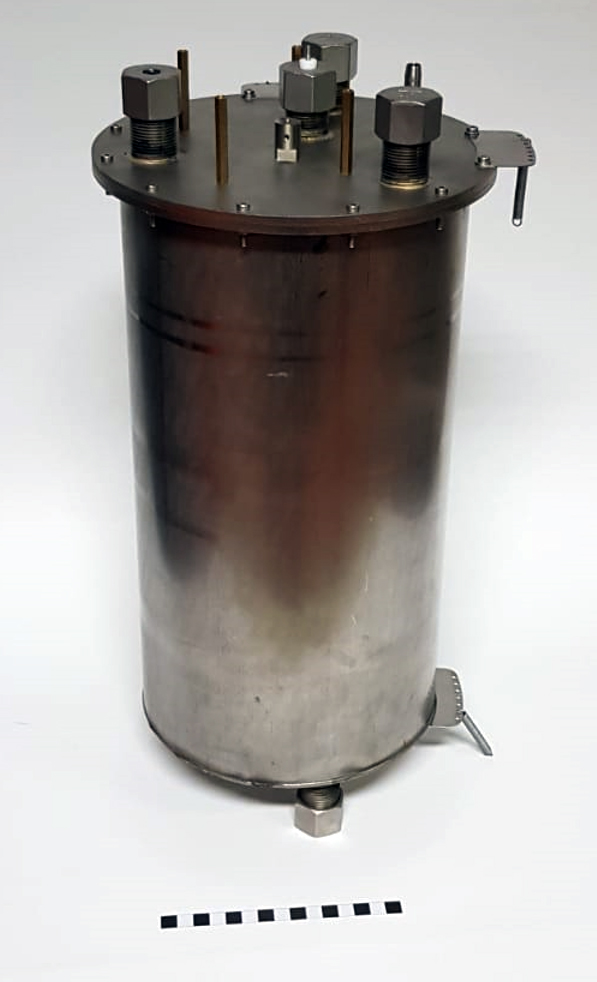}\hspace{6Ex}
    \includegraphics[height=0.42\linewidth]{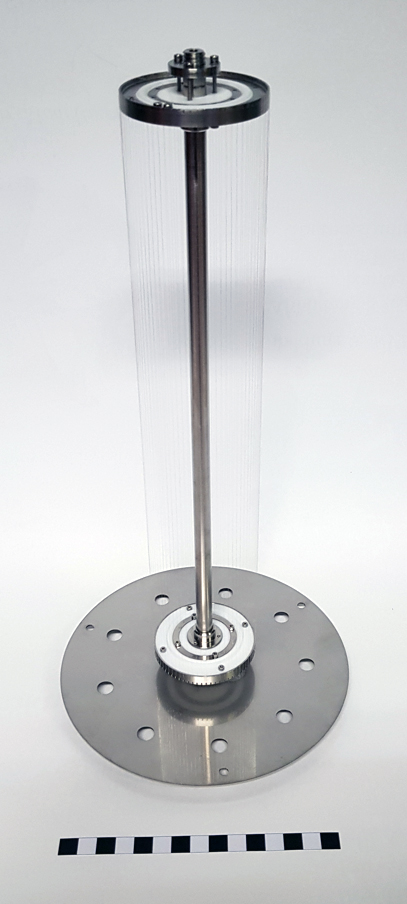}\hspace{6Ex}
    \includegraphics[height=0.42\linewidth]{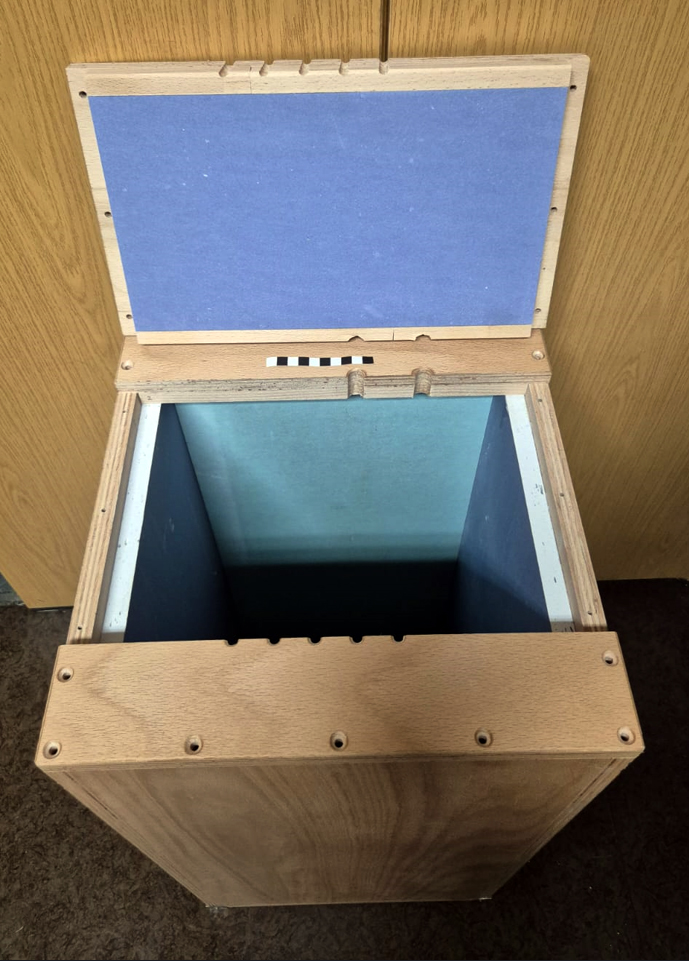}
    \caption{CIPIC II, its multi-wire cathode and the housing of the chamber}
    \label{fig:CIPIC}
\end{figure}

The ICs used for radon measurement to date follow design principle \textbf{I}, see \cite{Rottger_etal1998,LinzmaierDiss2013,Gavrilyuk_etal2011,Tianli_etmany2021,PEP2021,PEP2023}.
The desired equal path length for all ions is achieved, e.g. in \cite{LinzmaierDiss2013}, by a capacitor design in which the electrodes consist of parallel rods arranged at equal distances and spiraling around a longitudinal axis. The air to be analyzed is drawn in perpendicular to the longitudinal axis (see also Fig.\,3 in \cite{TeMe}).
The distances between the electrodes are typically $1\dots 3\,$cm, and the corresponding rod diameter is $0.5\dots 2\,$mm. As a result, the electric field between the electrodes is almost constant, so that the amplitude of the signals (theoretically) depends only on the energy of the $\alpha$ particles.

A disadvantage of this chamber design is the many obstacles along the $\alpha$ particles' trajectories. Depending on their energy, the range of $\alpha$ particles in air is approximately 4-8 cm \cite{Kilian:95765}. Due to the arrangement of the (relatively thick) electrodes, a large number of the $\alpha$ particles collide before they can convert their energy into ionized charge carriers. The result is a large number of low-amplitude electrical signals that cannot be assigned to $^{222}\mathrm{Rn}$ or its decay products. In \cite{Rottger_etal1998}, different design variants were investigated, and maximum resolutions (FWHM) of 10\% were achieved. A better-designed IC is described in \cite{Kuzminov2003}, where a resolution of 5\% is achieved.

Another disadvantage is that vibrations, e.g. due to room or impact noise, change the distance between the electrodes of the charged capacitor and thus its capacitance. This leads to voltage fluctuations, which interfere with the signal. Since the electrical pulse spectrum, the mechanical natural frequencies of the chamber, and the electrical interference from the 50\,Hz power grid are in the same range, this effect is considerable. Commercial IC use has  not yet been established therefore.

For this reason, the design principle \textbf{II} is pursued here, and a cylindrical, impulse-proportional ionization chamber (CIPIC) is constructed, cf. \cite {Gavrilyuk_etal2015}.
Two coaxial cylindrical tubes form the anodes of the capacitor, and the cathode is formed by thin wires arranged cylindrically. This construction avoids obstacles and allows for an undisturbed flight for the vast majority of $\alpha$ decays.

When an $\alpha$ particle travels undisturbed, the number of ions formed is proportional to the accumulated charge. The voltage-time area of the signal curve is therefore proportional to the energy of the $\alpha$ particle and can be separated and evaluated spectroscopically. For example, it is 5.49 MeV for $^{222}\mathrm{Rn}$, 6 MeV for $^{218}\mathrm{Po}$, and 7.68 MeV for $^{214}\mathrm{Po}$. 
Due to the trajectory's uncertain orientation relative to the IC geometry, the maximum signal height is highly variable at the same energy.

The capacitance \( C \) of a double cylinder capacitor is calculated as
\begin{align}\label{eq:Czyl}
  C &= C_{21} +  C_{32} = 2\pi \varepsilon_0 \varepsilon_r  \left( \frac{l}{\ln(D_2/D_1)} +\frac{l}{ \ln (D_3/D_2)} \right)
\end{align}
where \( \varepsilon_0 = 8.85 \times 10^{-12} \, \text{F/m} \) is the permittivity of vacuum and \( \varepsilon_r \) is the relative permittivity
of air; \( l \) is the length of the cylinders and \( D_1 \), \( D_2 \) and \( D_3 \) are the diameters of the inner, middle and outer electrodes.

Selecting these diameters carefully  can minimize the influence of mechanical disturbances. If the capacitance of the outer capacitor equals that of the inner one, $C_{32} = C_{21}$, then a vibrating wire  $D_2$ moves in such a way that the distances and thus the partial capacitances change annihilate.
Theoretically, a complete compensation of the capacitance changes is possible. From \reff{eq:Czyl}, we determine the condition
\begin{align}\label{eq:D2}
  D_2 = \sqrt{D_3\cdot D_1} \,.
\end{align}
Practical design considerations forced us to use different dimensions. Our two CIPICs illustrated in Figures\,\ref{fig:CIPICcut} and \ref{fig:CIPIC} and analyzed below are made of stainless steel. The cylindrical multi-wire cathode consists of 72 wires with a diameter of $50\,\upmu$m. The other dimensions and the effective chamber volume are summarized in Table 1.

As an additional measure for vibration isolation, the ICs were installed in a sturdy wooden box, see Fig.\,\ref{fig:CIPIC} (right). The interior of the box is lined with a layer of 18 mm thick plasterboard and mounted on specially designed elastomer buffers made of Regufoam \cite{BSW} with a relative volume of 0.15. Taken together, these measures result in excellent suppression of room and impact noise.

\begin{table}
\centering
\begin{tabular}{|l|c|c|c|c|r|} 
\hline
& length [mm] &   $D_1$ [mm] &  $D_2$ [mm] &  $D_3$ [mm]& volume [l] \\
\hline
CIPIC I &  370 & 168 & 64 & 10 & 7.7 \\
\hline
CIPIC II & 420 &  168 & 64 & 10 & 8.8\\
\hline
\end{tabular}
\caption{Dimensions of the two double-cylindrical CIPIC constructions}
\end{table}

\section{Data collection and processing}\label{sec:Elektrik}

To determine the ${^{222}\mathrm{Rn}}$ concentration with the CIPIC, the accumulated voltage signal is detected and evaluated spectroscopically.

\subsection{Measurement setup}
The ions are collected at the cathode at $D_2$, where they generate a positive charge. This results in a positive voltage of up to 100\,$\upmu$V. This signal is detected, amplified, and further evaluated using a sensitive spectroscopy amplifier.  To suppress electromagnetic interference, the system is surrounded by an electrically conductive shield that is grounded with low resistance.
\begin{figure}
    \centering
    \includegraphics[width=0.3\linewidth]{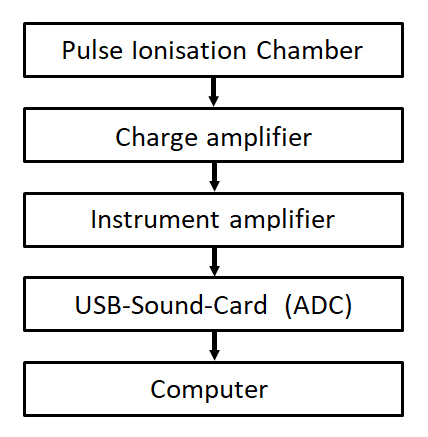}
    \caption{Signal path from detector to digital acquisition}
    \label{fig:SignalPath}
\end{figure}

The measuring chain has five steps and is shown in Fig.~\ref{fig:SignalPath}.

The CIPIC has an intrinsic capacitance of approximately 100\,pF. All insulators were made of polytetrafluoroethylene to ensure the highest possible insulation resistance (approx. 10 T$\Omega$). The spectroscopy amplifier largely corresponds to the design described in \cite{LinzmaierDiss2013}. A low-noise FET of type BF862 amplifies the resulting pulse. The dimensions were adapted to the existing conditions of the CIPIC through experiments, in particular the statistical mean of the charge-carrier collection time.
The discharge resistance was chosen to be 1\,G$\Omega$. Together with the input resistance of the spectroscopy amplifier, this results in a discharge time constant of $\tau = 3,5\,$ms. Given the variable charge-carrier collection time, this parametrization sufficiently meets the requirement
\begin{equation}\label{eq:Tau}
    \tau_\text{charge} = \tau_\text{discharge}
\end{equation}
The spectroscopy amplifier achieved a signal-to-noise ratio of 50:1. With the subsequent instrument amplifier (INA111), the reference mass was decoupled from the spectroscopy amplifier, and a low-impedance measurement signal was generated; the total gain is 2000. This signal is read into the computer using a standard USB sound card (ADC). For physical reasons, this measurement signal is unipolar. Data is recorded at a sampling rate of 44100 Hz, which fully complies with the Nyquist-Shannon theorem.

\subsection{Data processing }
Further evaluation and processing of the data is performed in Matlab \cite{Matlab2025}. For this purpose, the sampled data is stored as a wav file. After the selected measurement time (15-60 minutes) has elapsed, the wav file is sequentially searched for sections containing signal curves of $\alpha$ decays, and these sections are extracted. The result is a compact file containing the signal curves of all decays that have occurred. In the next step, overlapping signal regions are removed, and an offset compensation is applied to each pulse curve. A simple moving average filter with a window width of 1\,ms smooths the signal curves.
By summing the sampled input signal vector, the signal is integrated,  which provides the energy of the emitted $\alpha$ particle. These amounts are plotted in a histogram.

\begin{figure}
    \centering
    \includegraphics[height=0.3\linewidth]{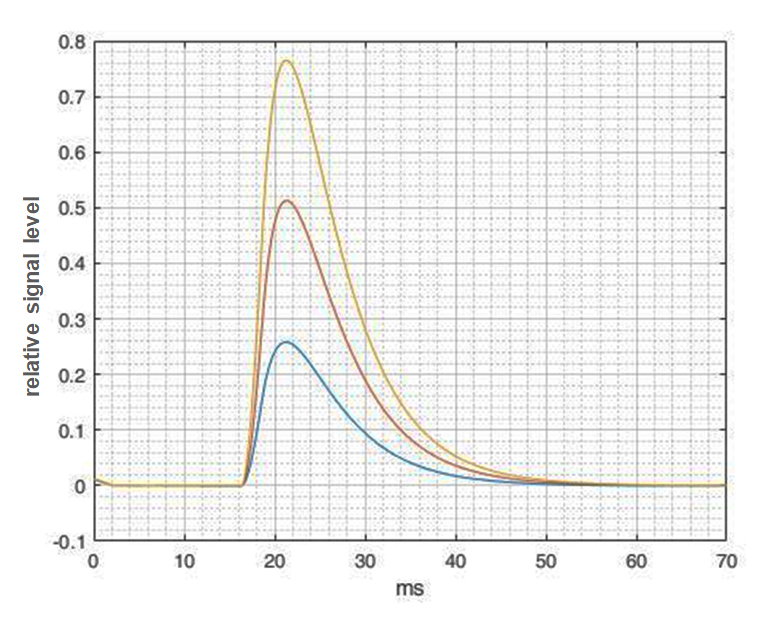}\hspace{1mm}
    \includegraphics[height=0.3\linewidth]{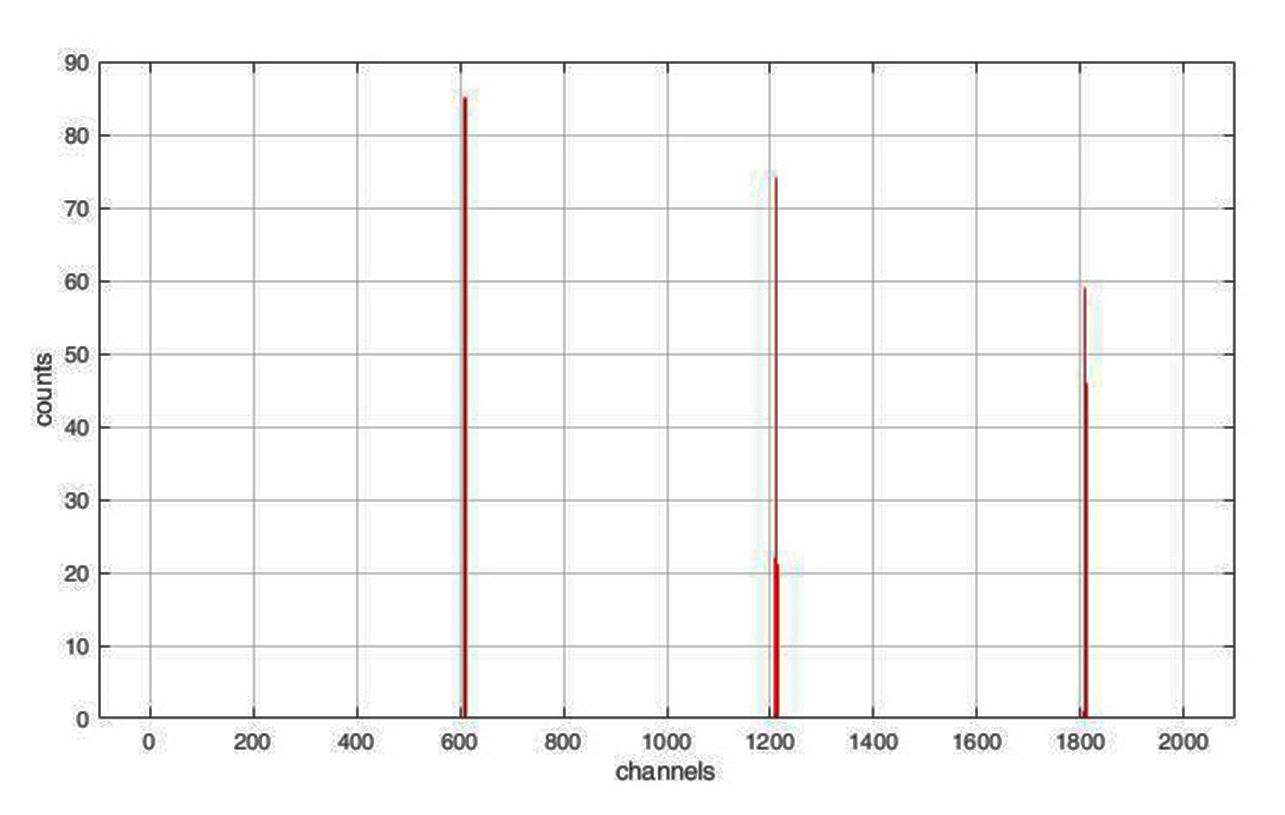}
    \caption{Artificially generated stress pulses (left) sampled with our signal path and the resulting histogram of energy {pulses} (right) in the ratio of 1:2:3}
    \label{fig:Signalpfadtest}
\end{figure}

\subsection{Verification}
To ensure that the signal chain shown in Fig.~\ref{fig:SignalPath} does not lead to distorted results due to error propagation, it was verified using artificial data. For this purpose, a test generator was developed that can generate synthetic $\alpha$ decay pulses. A rectangular pulse of defined width is generated, charging or discharging a capacitor and thereby allowing typical IC values to be set.

The signals in Fig.\,\ref{fig:CIPIC} (left) were generated by different charging times ($\tau_\text{charge}=10, 20, 30 \upmu$s at 100\,mA). The areas under the signal, i.e., the energy of the pulses,  have a ratio of 1:2:3. This is confirmed by the histogram in Fig.\,\ref{fig:CIPIC} (right), as the channel values are 609, 1212, and 1811 with deviations of $\pm 3$ channels, where 1 channel = 5\,keV. The systematic error in the signal path is therefore no more than 0.5\%.

\section{Results}\label{sec:Results}

\begin{figure}
    \centering
    \includegraphics[width=0.604\linewidth]{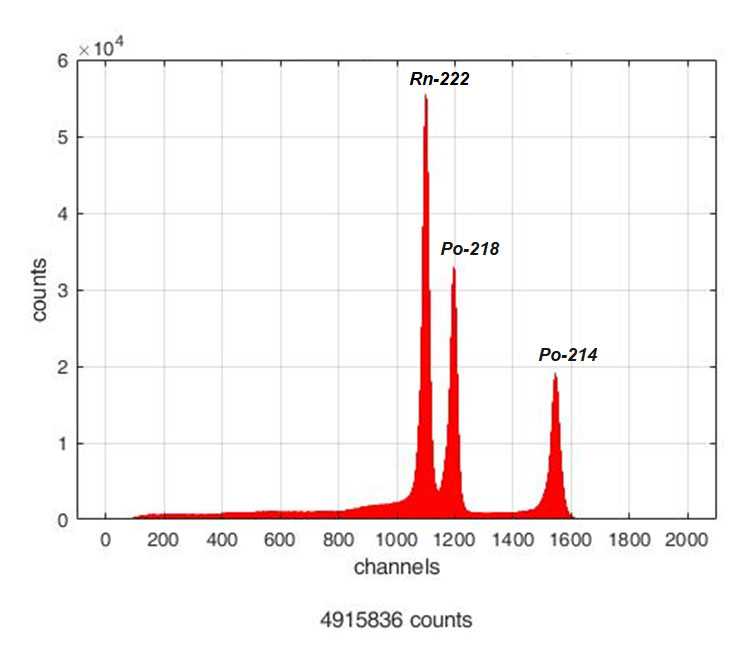}
    \caption{Histogram of a long-term measurement of  4915836 counts within 60 days at approx. $\mathrm{100\,Bq/m^3}$ }
    \label{fig:HistogrammCIPIC}
\end{figure}

\begin{figure}
    \centering
    \includegraphics[width=0.99\linewidth]{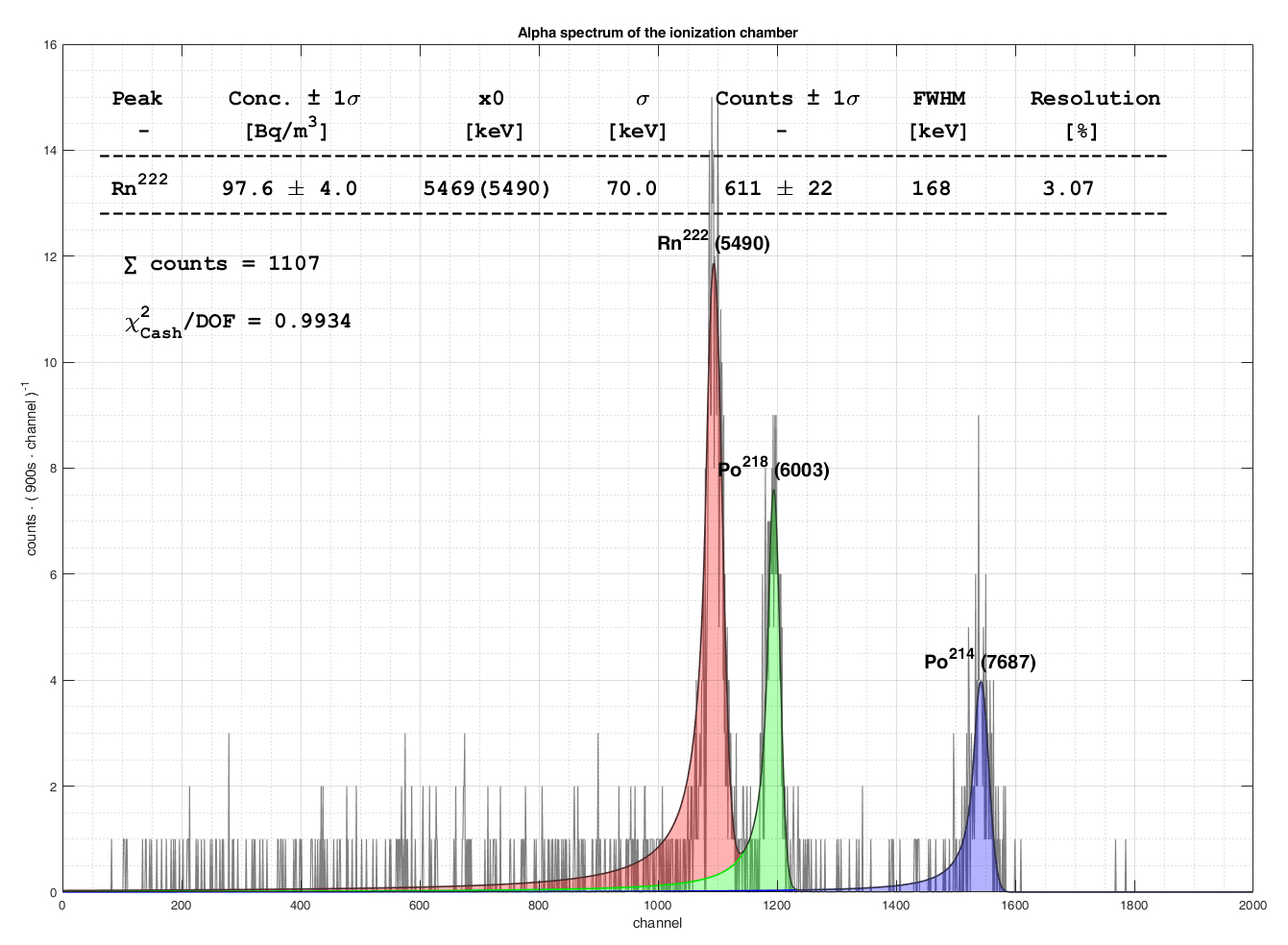}
    \caption{Histogram with evaluation of a 15-minute measurement cycle}
    \label{fig:CBF_CIPIC}
\end{figure}

To examine the measurement setup under realistic conditions, the constructed CIPICs were installed and put into operation in the solid mechanics laboratory at the University of Siegen.
The radon concentration in the ambient air here is in the range of $50 \dots 100\,\mathrm{Bq/m^3}$, which is typical for low-contaminated basement rooms.
Measurements were taken here over a period of several months; Fig.\,\ref{fig:HistogrammCIPIC} shows the results of CIPIC I.
The pronounced peaks distinctly indicate ${^{222}\mathrm{Rn}}$, ${^{218}\mathrm{Po}}$ and ${^{214}\mathrm{Po}}$ with a resolution of 2-3\% FWHM. Due to the high statistics, the CIPIC transfer function (instrument function) is clearly recognizable.

We additionally observed linearity of the CIPIC. With a coefficient of determination of 0.9975, the detected charges in the CIPIC correspond to the actual energy of the $\alpha$ decays in the relevant energy range from 5 to 8 MeV almost perfectly.

The ${^{222}\mathrm{Rn}}$ concentration is calculated by evaluating the histogram. The areas under the individual peaks correspond to the number of decay events. They are approximated using the crystal ball function (CBF). This fit is based on the crystal ball probability density function, which consists of a Gaussian core and a power-law tail. The power law is activated below a threshold, so the CBF has four free parameters to be fitted to the histogram data.

The CBF closely matches the actual instrument transfer function. To determine the area, a fit function consisting of three curves was created, which structurally corresponds to three CBFs. Since the entries per measurement channel are often less than 5 for short measurement times, the maximum likelihood method was used for fitting.
 
The activity of the air being examined can be calculated from the measurement time, the effective chamber volume, and the IC's determined efficiency. Thus, the CIPIC I detector with a volume of $7.7$l can measure an activity of 50\,Bq in the room air within 30\,min with a maximum uncertainty of 5\%. With the same accuracy, 5\,Bq can be measured within 5\,hours.

One of the periodic short measurements of the CIPIC 1 in continuous operation is shown as an example in Fig.\,\ref{fig:CBF_CIPIC}. With 611 decays assigned to the ${^{222}\mathrm{Rn}}$ peak, this yields an efficiency of 90\% and a resolution (relative FWHM) of 3.07\%. This is exceptionally good.

These real-time results are confirmed by offline calibration against the German SI standard at the Physikalisch-Technische Bundesanstalt (PTB), see \cite{TeMe}. Compared to an IC from the PTB based on design principle \textbf{I} \cite{LinzmaierDiss2013}, the CIPIC I performs significantly better.
The dispersion of the CIPIC measurement error has a Gaussian shape with an FWHM of only about $4\,\mathrm{Bq/m^3}$, which is an excellent result given an activity of less than $50\,\mathrm{Bq/m^3}$. During calibration, a standard uncertainty of $\pm 5\mathrm{Bq/m^3}$ was achieved within 1800\,s.

\begin{figure}
    \centering
    \includegraphics[width=0.75\linewidth]{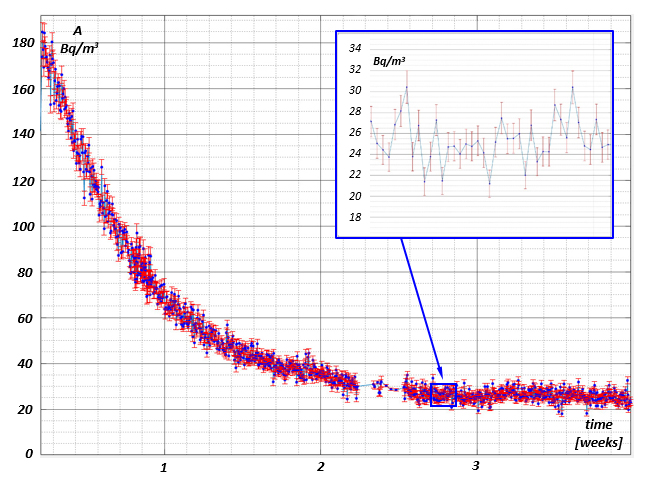}
    \caption{Decay curve of ${^{222}\mathrm{Rn}}$  activity over four weeks in closed CIPIC I} 
    \label{fig:CIPIC_Verlauf}
\end{figure}

Furthermore, a long-term measurement was carried out over several weeks, starting with a ${^{222}\mathrm{Rn}}$ activity of $200\,\mathrm{Bq/m^3}$, which was generated by means of a ${^{226}\mathrm{Ra}}$ emission source. After closing the CIPIC, the activity decays. 
The exponential decay of ${^{222}\mathrm{Rn}}$ was measured every 15 minutes and is recorded in Fig.\,\ref{fig:CIPIC_Verlauf}.
The half-life of 4 days is clearly visible. In the inset, it can be seen that despite the short measurement time of 15 minutes and the low concentration of $24\,\mathrm{Bq/m^3}$, the statistical uncertainty is only $\pm2\,\mathrm{Bq/m^3}$, i.e. less than 10\,\%.
To our knowledge, no other ${^{222}\mathrm{Rn}}$-measurement system can achieve this.

\section{Summary}\label{sec:Summary}

This paper presents an impulse-proportional IC for detecting $\alpha$ decays in ambient air, suitable for direct measurement of low radon concentrations and achieving spectroscopic resolutions of 2-3\%.

We can achieve this accuracy through two novel principles. On the one hand, the IC's double-cylindrical, coaxial design allows efficient, near-complete detection of $\alpha$ radiation by avoiding obstacles in the trajectories of ionized particles. The charges generated in this process are energy-proportional and can be evaluated with an efficiency of 90\%
On the other hand, we developed a customized measurement method for spectroscopic evaluation to discriminate between the proportions of {${^{222}\mathrm{Rn}}$, ${^{218}\mathrm{Po}}$ and ${^{214}\mathrm{Po}}$} and extract their concentrations.

Attention was paid to the energy resolution of the detection system by suppressing the effects of acoustic and vibration noise on the detector's operation.

The high spectral resolution of the developed IC  allows measurements of very low ${^{222}\mathrm{Rn}}$ activity concentrations. With a relatively large chamber volume of 7.7 liters,
 the IC enables measurement uncertainties of less than 5\% in 15-minute measurement times, e.g., in air with $50\,\mathrm{Bq/m^3}$ radon activity. The detector is calibrated and traceable to the SI via the German national standard for $\alpha$-emitting activity sources based on ${^{226}\mathrm{Ra}}$.

Future investigations will focus on CIPICs with larger and smaller volumes, for specific indoor or outdoor use, and on their measurement characteristics.
In addition, the materials used in the construction of the detector will be examined for contamination with ${^{226}\mathrm{Ra}}$ and thus for interference from ${^{222}\mathrm{Rn}}$.


\begin{thebibliography}{10}

\bibitem{BfS2024}
\protect{BfS}.
\newblock Legal regulations for the protection against radon, 2024.
\newblock \url{www.bfs.de/EN/topics/ion/environment/radon/regulations/law.html}, last accessed: 2025-03-05.

\bibitem{EU}
European Commission.
\newblock \url{https://energy.ec.europa.eu/topics/nuclear-energy/radiation-protection_en}, 
\newblock Radiation protection legislation, last accessed: 2025-03-05.

\bibitem{Chambers2018}
S.D. Chambers, S.~Preunkert, R.~Weller, S.-B. Hong, R.S. Humphries, L.~Tositti, H.~Angot, M.~Legrand, A.G. Williams, A.D. Griffiths, J.~Crawford, J.~Simmons, T.J. Choi, P.B. Krummel, S.~Molloy, Z.~Loh, I.~Galbally, S.~Wilson, O.~Magand, F.~Sprovieri, N.~Pirrone, and A.~Dommergue.
\newblock Characterizing atmospheric transport pathways to antarctica and the remote southern ocean using Radon-222.
\newblock {\em Frontiers in Earth Science}, 6, 2018.

\bibitem{LukasWoodward1964}
HF~Lucas~Jr and DA~Woodward.
\newblock Effect of long decay chains on the counting statistics in the analysis of Radium224 and Radon222.
\newblock {\em Journal of Applied Physics}, 35(2):452--456, 1964.

\bibitem{WilliamsChambers2016_HistoryColl}
A.G. Williams and S.D. Chambers.
\newblock {A History Of Radon Measurements At Cape Grim}.
\newblock \emph{ANSTO},
\newblock \protect{Australian} Nuclear Science and Technology Organisation, Kirrawee DC, NSW1, 2016. 

\bibitem{Sarad2024}
Sarad GmbH.
\newblock Radon \& Thoron,
\newblock \url{https://www.sarad.de/cms/media/docs/applikation/Vergleich_Messprinzipien.pdf}, last accessed: 2025-03-05.

\bibitem{Pronost_etal2019}
G.~Pronost, M.~Ikeda, T.~Nakamura, H.~Sekiya, and S.~Tasaka.
\newblock Development of new radon monitoring systems in the Kamioka mine.
\newblock {\em Progress of Theoretical and Experimental Physics}, 2019(7):079201, 07 2019.

\bibitem{Kuzminov2003}
V.V. Kuzminov.
\newblock Ion-pulse ionization chamber for direct measurement of radon concentration in the air.
\newblock {\em Phys. Atom. Nuclei}, 66:462--465, 2003.

\bibitem{Gavrilyuk_etal2011}
Y.M. Gavrilyuk, A.M. Gangapshev, V.V. Kuzminov, S.I.Panasenko, and S.S. Ratkevich.
\newblock Monitoring the \protect{$^{222}$Rn} concentration in the air of low-background laboratories by means of an ion-pulse ionization chamber.
\newblock {\em Bull. Russ. Acad. Sci. Phys.}, 75:547--551, 2011.

\bibitem{Kozlov_etmany2020}
A.~Kozlov, D.~Chernyak, Y.~Takemoto, K.~Fushimi, K.~Imagawa, K.~Yasuda, H.~Ejiri, R.~Hazama, H.~Ikeda, K.~Inoue, S.~Yoshida, R.A. Etezov, Yu.~M. Gavrilyuk, V.V. Kazalov, V.V. Kuzminov, and S.I. Panasenko.
\newblock Detectors for direct dark matter search at Kamland.
\newblock {\em Nuclear Instruments and Methods in Physics Research Section A: Accelerators, Spectrometers, Detectors and Associated Equipment}, 958:162239, 2020.

\bibitem{Etezov_etal2022}
R.A. Etezov, Yu.M. Gavrilyuk, A.M. Gangapshev, A.M. Gezhaev, V.V. Kazalov, A.Kh. Khokonov, V.V. Kuzminov, S.I. Panasenko, and S.S. Ratkevich.
\newblock \protect{$^{222}$Rn} content variations at ground and underground conditions.
\newblock arXiv 2110.15289, 2022.

\bibitem{Gavrilyuk_etal2015}
Yu.M. Gavrilyuk, A.M. Gangapshev, A.M. Gezhaev, R.A. Etezov, V.V. Kazalov, V.V. Kuzminov, S.I. Panasenko, S.S. Ratkevich, D.A. Tekueva, and S.P. Yakimenko.
\newblock High-resolution ion pulse ionization chamber with air filling for the \protect{$^{222}$Rn} decays detection.
\newblock {\em Nuclear Instruments and Methods in Physics Research Section A: Accelerators, Spectrometers, Detectors and Associated Equipment}, 801:27--33, 2015.

\bibitem{Rottger_etal1998}
Stefan R{\"o}ttger, A~Paul, A~Honig, and U~Keyser.
\newblock Vieldraht Impulsionisationskammern zur Pr{\"a}zisionsmessung der Radon-Aktivit{\"a}tskonzentration in Luft.
\newblock Technical Report, CM-P00066600, 1998.

\bibitem{LinzmaierDiss2013}
D.~Linzmaier.
\newblock {\em {Entwicklung einer Low-Level-Radon-Referenzkammer}}.
\newblock PhD thesis, Gottfried Wilhelm Leibniz Universität Hannover, 2013.

\bibitem{Tianli_etmany2021}
T. Qiu, M. Li, X. Wei, H. Yang, P. Ma, C. Lu, L. Duan, R. Hu, Z. He, J. Liang, and M. Zhang.
\newblock Ion pulse ionization chamber for online measurements of the radon activity concentration.
\newblock {\em Nuclear Techniques}, 44(4):040403, 2021.

\bibitem{PEP2021}
F.~Schneider M.~Schmidt, F.~Oerter.
\newblock Design and construction of a test facility for determining the activity and half-life of the noble gas Radon (in German).
\newblock Students project, Universität Siegen, 2021.

\bibitem{PEP2023}
P.~Beckmann M.~Geethan, J.~Heuel.
\newblock Design and construction of a portable Radon pulse ionization chamber CPIC (in German).
\newblock Students project, Universität Siegen, 2023.

\bibitem{TeMe}
Stefan Röttger, Ralf Nötzel, Anja Honig, Benoit Sabot, and Kerstin Weinberg.
\newblock Radon sensor networks for large buildings: balancing the trade-off between energy efficiency and health.
\newblock {\em tm - Technisches Messen}, 92(9-10):382--391, 2025.

\bibitem{Kilian:95765}
Ulrich Kilian and Christine Weber.
\newblock {\em {L}exikon der {P}hysik: {B}and 1. {A}a bis {D}e;}
\newblock Spektrum Akademischer Verlag GmbH, Heidelberg, Neckar, 2003.

\bibitem{BSW}
Berleburger~Schaumstoffwerke GmbH.
\newblock \protect{Regufoam} \protect{Schwingungstechnik} Technische \protect{Daten}, 2024.
\newblock Release 6, \url{http://www.bsw-schwingungstechnik.de}, last accessed: 2025-06-17.

\bibitem{Matlab2025}
Matlab, 2025.
\newblock \url{https://de.mathworks.com/products/matlab.html}, last accessed: 2025-03-05.

\end{thebibliography}


\section*{Acknowledgement}
The autors gratefully acknowledge the support of the 23IND07 RadonNET project 'Radon metrology: Sensor networks for large buildings and future cities' as well as the  support provided by the workshops of the University of Siegen.

\section*{Declaration of competing interest}
The authors declare that they have no competing financial interests or personal relationships that could have appeared to influence the work reported in this paper.
%

\printcredits

\section*{Data availability}
Data will be made available on reasonable request.

\section*{Funding}
The project 23IND07 RadonNET has received funding from the European Partnership on Metrology, co-financed from the European Union’s Horizon Europe Research and Innovation Programme and by the Participating States.

\end{document}